\documentclass[12pt, amsmath, amssymb, nofootinbib, sort&compress, showkeys, round, citeautoscript]{revtex4}

\setcitestyle{square}

\usepackage{graphics}
\usepackage{epsfig}
\usepackage{subfigure}  
\usepackage{bm}
\usepackage{booktabs}       
\usepackage{threeparttable} 

\makeatletter
\def\@biblabel#1{(#1)}
\makeatother

\begin{document}
\title{ Modified Becke-Johnson exchange potential: improved modeling of lead halides for solar cell applications}

\author{
Radi A. Jishi}

\email{rjishi@calstatela.edu;   Tel:323-343-2137}

\affiliation{ Department of Physics, California State University, Los Angeles, California, U.S.A}


\date{\today}

\begin{abstract}
We report first-principles calculations, within density functional theory, on the lead halide
compounds PbCl$_2$, PbBr$_2$, and CH$_3$NH$_3$PbBr$_{3-x}$Cl$_x$, taking into account spin-orbit
coupling. We show that, when the modified Becke-Johnson exchange potential is used with a suitable
choice of defining parameters, excellent agreement between calculations and experiment
is obtained. The computational model is then used to study the effect of replacing the methylammonium
cation in CH$_3$NH$_3$PbI$_3$ and CH$_3$NH$_3$PbBr$_3$ with either N$_2$H$_5^+$ or N$_2$H$_3^+$, which 
have slightly smaller ionic radii than methylammonium. We predict that a considerable
downshift in the values of the band gaps occurs with this replacement. The resulting compounds 
would extend optical absorption  down to the near-infrared  region, creating 
excellent light harvesters for solar cells.
\end{abstract}


\maketitle

\section{Introduction}\label{sec:1}

Organolead halide perovskites are attracting great interest, mainly because of their
photovoltaic applications. These compounds have the general formula ABX$_3$, where A is an organic 
cation, B is lead, and X is a halide ion. An example is methylammonium lead iodide (CH$_3$NH$_3$PbI$_3$,
abbreviated as MAPbI$_3$), where the MA ion is coordinated to 12 I ions, while Pb is octahedrally
coordinated to six I ions, with every two adjacent octahedra sharing a corner.

Following publication of the first report \cite{ref:Kojima} on the use of these 
compounds as light harvesters 
in photovoltaic cells, many experimental studies have tried to improve upon material
preparation methods  
 and enhance their solar-to-electric power conversion 
efficiency. \cite{ref:Etgar, ref:Ball, ref:Heo, ref:Kim, ref:Bi, ref:Cai, ref:Eperon1, ref:Laban, 
 ref:Stranks, ref:Lee, ref:Noh, ref:Burschka, ref:Liu} 
Theoretical studies have also been undertaken to explore the electronic 
properties of these compounds and to develop accurate models for computing the energy bands.
\cite{ref:Mosconi, ref:Wang, ref:GW, ref:Even1, ref:Even2, ref:Even3, ref:Feng, ref:Brivio,
 ref:Filippetti, ref:Jishi, ref:Motta}

MAPbI$_3$ has a band gap of 1.5-1.6 eV. \cite{ref:Noh, ref:Baikie} The possibility of
 tuning the band gap by various substitutions
has been considered. One approach is to replace I with Br or Cl, yielding MAPbBr$_3$ or MAPbCl$_3$,
with band gaps given by 2.22 eV and 3.17 eV, respectively.\cite{ref:Noh, ref:Comin, ref:Buin} 
Another approach is to replace the 
methylammonium cation with another cation of a different size. When CH$_3$NH$_3$ is replaced with 
FA = NH$_2$CHNH$_2$ (formamidinium), FAPbI$_3$ with a band gap of 1.43-1.48 eV is obtained.
\cite{ref:Pang, ref:Stoumpos1, ref:Stoumpos2, ref:Eperon2, ref:Koh}   
The lower band gap in FAPbI$_3$, compared to the one in MAPbI$_3$, leads to an increase in the material's optical
absorption range. Unfortunately, FAPbI$_3$ is unstable under ambient conditions. However, it has been
shown recently that the incorporation of MAPbBr$_3$ into FAPbI$_3$ stabilizes the perovskite phase
of FAPbI$_3$ and enhances the power conversion efficiency of the solar cell 
to 18 percent.\cite{ref:Jeon}

In addition to the extensive work demonstrating the applications of organometallic halides in solar cells, it 
has been shown that these materials  have optoelectronic applications in light-emitting 
diodes.\cite{ref:Tan, ref:Kim2} For such applications, it is desirable to have materials where the optical band gap 
can be tuned over a wide range of the visible spectrum. Recent studies \cite{ref:Comin} 
have shown that, in 
MAPbBr$_3$, it is easy to substitute Cl for Br, yielding air-stable MAPbBr$_{3-x}$Cl$_x$ with 
$x$=0-3. The simple cubic lattice structure of MAPbBr$_3$ is maintained as $x$ increases, and the 
lattice constant decreases linearly with increasing $x$; the reduction is 5\% at $x=3$. On the 
other hand, the band gap increases with increasing $x$, leading to a tunable band gap in the 400 
to 550 nm wavelength range.

The crystal structure of the organometallic halides depends on the radii of the constituent ions 
through the Goldschmidt tolerance factor $t$, which, for the compound ABX$_3$, is given by

\begin{equation}
 t = \frac{r_A + r_X}{\sqrt{2}(r_B + r_X)}
\end{equation}
where $r_A$, $r_B$, and $r_X$ are the ionic radii of the A, B, and X ions, respectively. 
This factor serves as a guide to predicting the crystal structure of ABX$_3$; for values of $t$ 
ranging from 0.9 to 1.0, an ideal cubic perovskite structure is favored. 

In calculating $t$, one is 
faced with the problem of what value to use for the ionic radius of the organic cation A; different 
values result from different methods of calculation. Amat et al.\cite{ref:Amat} obtained an ionic radius of 
2.70 \AA $ $ for CH$_3$NH$_3^+$ by calculating the volume inside  a contour of 0.001 electrons/bohr$^3$ 
density. Kieslich et al.\cite{ref:Kieslich} considered a hard sphere model in which the cation rotates freely about
its center of mass. The ionic radius of CH$_3$NH$_3^+$ is then taken to be the distance from the cation's
center of mass  to nitrogen, plus 1.46 \AA \ (the ionic radius of nitrogen). This 
method yields a value of 2.17 \AA $ $ for the ionic radius of the methylammonium ion. 

We can set some 
bounds on the value of the ionic radius of CH$_3$NH$_3^+$ by noting that both MAPbBr$_3$ and 
MAPbCl$_3$ adopt an ideal cubic perovskite structure. Taking the ionic radii of Pb, Br, and Cl to be
1.19 \AA, 1.96 \AA, and 1.81 \AA, respectively, we find that for the tolerance factor $t$ to lie 
between 0.9 and 1.0, the ionic radius of CH$_3$NH$_3^+$ should lie between 2.04 \AA $ $ and 2.50 \AA,
These values should not be considered  strict limits; they are reasonable estimates. We should
note that, in organolead halide compounds, the methylammonium ion does not rotate freely
about its center of mass. There is disorder, manifested by the existence of 12 equivalent
positions for C and N. In all these positions, the midpoint of the C-N bond is always at, or
extremely close to, the center of the cubic unit cell.\cite{ref:Mashiyama} Thus, we may estimate the ionic radius of
CH$_3$NH$_3^+$ as one half the C-N bond length plus the ionic radius of nitrogen. Optimizing the 
structure of the methylammonium cation by using the 6-31G** basis set of gaussian orbitals 
and the B3LYP exchange potential \cite{ref:B3LYP} 
as implemented within Gaussian 09, \cite{ref:Gaussian} 
 the ionic radius of the cation is found to be 2.23 \AA. The resulting Goldschmidt tolerance
factors are 0.952, 0.941, and 0.924 for MAPbCl$_3$, MAPbBr$_3$, and MAPbI$_3$, respectively. These values
are consistent with the fact that these compounds adopt a perovskite structure.

The electronic properties of  organometallic halides depend on various factors which can be controlled
experimentally. These factors include  lattice constants, which can be varied by applying 
external pressure or internal chemical pressure; the type of halide ion, controlled by chemical 
substitution; and the type of  organic ion. To obtain meaningful results, it is important to use a computational model 
which accurately describes the known electronic properties of these compounds and which can  
 predict the effect of these variables upon them. Density functional
theory (DFT) in the Kohn-Sham formulation \cite{ref:Kohn} is the most widely used method. 
Here, the exchange potential is approximated by a 
functional of the electronic density, with the most common approximations being the local density 
approximation (LDA) \cite{ref:Kohn} and the generalized gradient approximation (GGA). \cite{ref:Perdew}
 Although the ground
state is well described by LDA and GGA, these approximations fail to account for excited-state 
properties. In many semiconductors the values of the band gaps are severely underestimated. Improved
values for the band gaps are obtained by using the GW method.\cite{ref:Bechstedt}  The usefulness of this method,
however, is hampered by its high computational cost.

A different exchange potential, introduced by Becke and Johnson, \cite{ref:Becke06} was recently modified by Tran and 
Blaha.\cite{ref:Tran09} The  modified Becke-Johnson (mBJ) potential is given by
\begin{equation}
  V_{mBJ}(\mathbf{r})=cV_x^{BR}(\mathbf{r})+(3c-2)\frac{1}{\pi}\sqrt{\frac{5}{12}}~[2t(\mathbf{r})/\rho(\mathbf{r})]^{1/2}
\label{eq:tbmbjpot}
\end{equation} 
where  $ V_x^{BR}(\mathbf{r}) $ 
is the Becke-Roussel exchange potential,\cite{ref:Becke89}  $ \rho(\mathbf{r}) $ is the electron 
density, and $ t(\mathbf{r}) $ is the Kohn-Sham  kinetic 
energy density. In the above equation,
\begin{equation}
 c = A + B \sqrt{g}
\end{equation}
where $g$ is the average of  $|\bm{\nabla}\rho/\rho|$ over the volume of the unit cell, and $A$ and 
$B$ are parameters adjusted to produce the best fit to the experimental values of the 
semiconductor band gaps.

We have recently shown that, upon using the modified Becke-Johnson exchange
potential with $A = 0.4$ and $B = 1.0$ bohr$^{1/2}$, the calculated band gaps of MAPbI$_3$, MAPbBr$_3$, RbPbI$_3$,
and CsPbX$_3$ ( X = Cl, Br, I) are in excellent agreement with  experimental values. \cite{ref:Jishi}
In this work, we show that applying this method to the lead halide compounds PbCl$_2$, 
PbBr$_2$, and MAPbBr$_{3-x}$Cl$_x$ for $x = 1, 2$, and $3$ produces accurate values for the band gaps.
We then use this method to demonstrate that a small reduction in the lattice constants of 
MAPbBr$_3$ and MAPbI$_3$ produces a considerable downshift in the band gaps. Reduction in the 
lattice constants can be achieved by replacing  CH$_3$NH$_3^+$ with slightly smaller cations,
such as N$_2$H$_5^+$ and N$_2$H$_3^+$.

\section{Methods}\label{sec:2}

Total energy calculations are carried out using the all-electron, full potential, linearized
 augmented plane wave (FP-LAPW) method as implemented in the WIEN2k code.\cite{ref:Blaha} 
In this method, space is divided into two regions.  One region consists of the interior of 
 non-overlapping muffin-tin spheres centered at the atom sites. The rest of the space 
(the interstitial) constitutes the other region. In all the calculations reported in this work,
the radii of the muffin-tin spheres are 2.1$a_0$ for Pb, Cl, Br, and I, where $a_0$ is 
the Bohr radius. On the other hand, the radii for C, N, and H are chosen such that the 
muffin-tin spheres on adjacent atoms almost touch. The electronic wave function is expanded
in terms of a set of basis functions which take different forms in the two regions
mentioned above. Inside the muffin-tin spheres, the basis functions are atomic-like functions
which are expanded in terms of spherical harmonics up to $l_{max}=10$. In the interstitial
region, they are plane waves with a maximum wave vector $K_{max}$. Each plane wave is augmented
by one atomic-like function in each muffin-tin sphere. Usually, $K_{max}$ is chosen such that
 $R_{mt}K_{max} =$ 6-9, where $R_{mt}$ is the radius of the smallest muffin-tin sphere in the 
unit cell. However, due to the very smallness of the muffin-tin radius of hydrogen ($R_H =$ 0.65-0.70 $a_0$),
we set $R_H K_{max} = 3.0$. For orthorhombic and tetragonal systems, 
a 4x4x4 Monkhorst-Pack grid \cite{ref:Monkhorst} was used for sampling the Brillouin zone, while 
an 8x8x8 grid was chosen for cubic systems.

The charge density is Fourier-expanded up to a maximum wave vector $G_{max}= 20 a_0^{-1}$. In GGA 
calculations, PBEsol functional \cite{ref:PBEsol} is used for the exchange potential. This method
is well-suited for optimizing atomic positions and lattice constants. In the self-consistent
calculations, the total energy and charge were converged to within 0.1 mRy and 0.001 e, respectively.
In calculations which employ the modified Becke-Johnson exchange potential, we choose $A=0.4$ and
$B=1.0$ bohr$^{1/2}$, where $A$ and $B$ are the parameters which appear in Eq.(3).

\section{Results and Discussion}\label{sec:3}

We begin by carrying out DFT calculations on PbCl$_2$, PbBr$_2$, and MAPbBr$_{3-x}$Cl$_x$, for 
$x = 0,1, 2$, and $3$. At room temperature, PbCl$_2$ and PbBr$_2$ adopt an orthorhombic crystal structure
\cite{ref:Wyckoff} with space group Pbnm, while MAPbBr$_{3-x}$Cl$_x$ crystals have a cubic unit cell
\cite{ref:Noh, ref:Comin}. The experimental values of the lattice constants of these crystals are 
given in Table 1.

\begin{table}[ht]
  \caption{Crystal structure and lattice constants for some of the compounds studied in this work.}
      \begin{tabular}{ccc}
	\hline
	\hline
	Compound & Structure & Lattice constants (\AA)\\
	\hline
	PbCl$_2$                    & Orthorhombic	& a=9.03, b=7.608, c=4.525				\\
	PbBr$_2$                    & Orthorhombic	& a=9.466, b=8.068, c=4.767	\\
	MAPbBr$_3$                  & Cubic     	& a=5.933	\\
	MAPbBr$_2$Cl                & Cubic	        & a=5.88	\\
	MAPbBrCl$_2$		    & Cubic		& a=5.78				\\
        MAPbCl$_3$                  & Cubic             & a=5.71       \\
	\hline
	\hline
      \end{tabular}
      \label{tab:1}
\end{table}

In DFT calculations, it is important to take into account spin-orbit coupling, mainly 
because of the presence of Pb. Our results are summarized in Table 2. In all the compounds under 
consideration, we find that GGA+SOC severely underestimates the values of the band gaps. On the 
other hand, our present method (mBJ+SOC, with $A=0.4$ and $B=1.0$ bohr$^{1/2}$ in Eq. (3)) 
yields values for the band gaps which
are in excellent agreement with experiment. These results, along with previous calculations
\cite{ref:Jishi} on other lead halide compounds, give us confidence in the ability of DFT
combined with the mBJ exchange potential to  accurately predict the band gaps in all lead halide compounds.



\begin{table*}[t]  
  \caption{Calculated and experimental band gaps, in eV, for the compounds that are studied in this work.}
      \begin{tabular}{cccccccc}
	\hline
	\hline 
        Compound                   & GGA+SOC 	& mBJ+SOC	& Experiment		\\
	\hline
	PbCl$_2$                    & 3.18      & 5.13		& 5.38, \cite{ref:Plekhanov} 4.86 \cite{ref:Zaldo}	\\
	PbBr$_2$                    & 2.46	& 4.19  	& 4.23, \cite{ref:Plekhanov2} 4.1  \cite{ref:Iwanaga}  \\
	MAPbBr$_3$                  & 0.45	& 2.23  	& 2.28 \cite{ref:Noh}	  	\\
	MAPbBr$_2$Cl                & 0.46	& 2.42  	& 2.65 	\cite{ref:Comin}  	\\
	MAPbBrCl$_2$                & 0.57	& 2.76  	& 2.90 	\cite{ref:Comin}  	\\
	MAPbCl$_3$                  & 1.0	& 3.22  	& 3.17 	 \cite{ref:Comin}	\\
	\hline
	\hline
      \end{tabular}
      \label{tab:2}
\end{table*}

In Fig.1 we present the calculated density of states in PbBr$_2$. The figure shows that the 
lower-lying conduction bands are derived from Pb  orbitals, while  
 both Pb  and Br  orbitals contribute to the highest valence band. The situation is similar in PbCl$_2$ and
MAPbBr$_{3-x}$Cl$_x$; the low conduction bands are derived mainly from Pb  orbitals while the 
valence band is composed of Pb  and halide  orbitals.

\begin{figure}[htbp]
   \includegraphics[height=8.5cm]{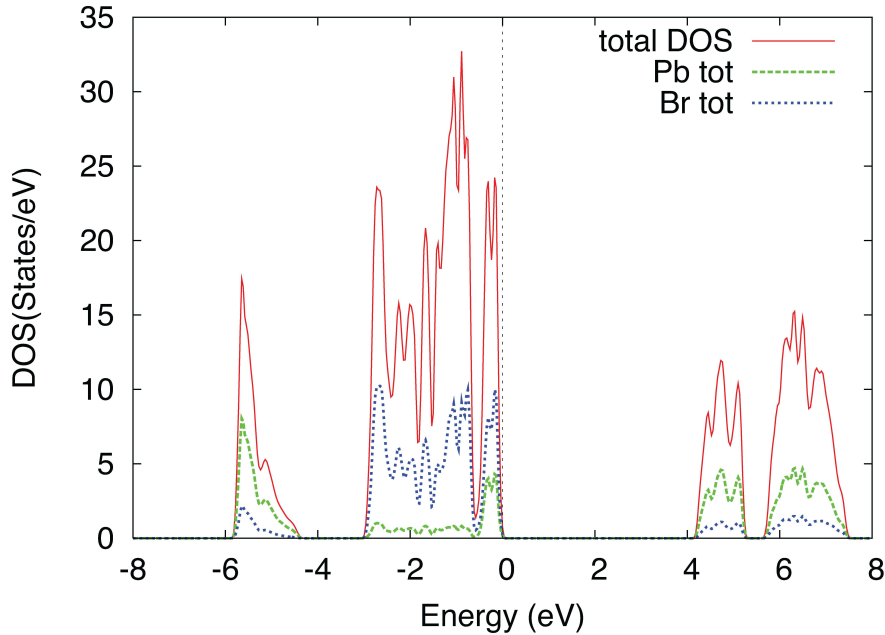}
  \caption{  Density of states (DOS) of PbBr$_2$,    
obtained using the mBJ potential, with $A$ and $B$ in Eq.(3) given by 0.4 and 1.0 bohr$^{1/2}$, respectively, and 
taking into account the effect of spin-orbit coupling.
  \label{fig:1}}
\end{figure}

We now apply this method to  study  the variation of the band gap in MAPbBr$_3$ and 
MAPbI$_3$ with the reduction of the lattice constants. Our results indicate that the band gap in 
these materials is very sensitive to lattice constant variation. At the experimental values of 
the lattice constants, our calculations give a band gap of $E_g = 1.54$ eV in MAPbI$_3$ and $E_g=2.23$ eV
in MAPbBr$_3$. For a 1\% reduction in the lattice constants, we obtain 
$E_g = 1.31$ eV in MAPbI$_3$ and $E_g=1.95$ eV in MAPbBr$_3$, while for a 2\% reduction, we obtain
$E_g = 1.17$ eV in MAPbI$_3$ and $E_g=1.75$ eV in MAPbBr$_3$. In arriving at these results, we have 
assumed that, with reduced lattice constants, MAPbI$_3$ maintains its body-centered tetragonal
structure, and MAPbBr$_3$ its simple cubic structure.

A reduction in the lattice constants of organometallic halides can be achieved by applying an external
pressure or by replacing CH$_3$NH$_3^+$ with a cation of a slightly smaller ionic radius. We consider
two cations: N$_2$H$_5^+$ (diazanium) and N$_2$H$_3^+$ (diazenium). In each case, the ionic radius is 
taken to be equal to one-half the N-N bond length plus the ionic radius of nitrogen, as discussed
in the introduction. Using coupled cluster theory with a perturbative treatment of the triple excitations,
Matus et al.\cite{ref:Matus} calculated the N-N bond lengths in  N$_2$H$_5^+$ and N$_2$H$_3^+$
to be 1.46 \AA \ and 1.24 \AA, respectively. Using these values, we obtain 2.19 \AA \ and
2.08 \AA \  for the ionic radii of N$_2$H$_5^+$  and N$_2$H$_3^+$, respectively. These values
are only slightly smaller than the corresponding value for CH$_3$NH$_3^+$, estimated by the 
same method to be 2.23 \AA. The Goldschmidt tolerance factors for N$_2$H$_5$PbBr$_3$ and
N$_2$H$_5$PbI$_3$ are 0.932 and 0.916, respectively, while for N$_2$H$_3$PbBr$_3$ and
N$_2$H$_3$PbI$_3$ they are  0.907 and 0.893, respectively. Thus, the replacement
of CH$_3$NH$_3$ with N$_2$H$_5$ or N$_2$H$_3$ leads to only a small change in the 
tolerance factor. Hence, we assume that N$_2$H$_5$PbBr$_3$ and N$_2$H$_3$PbBr$_3$ 
will have a cubic unit cell, similar to MAPbBr$_3$, and
that N$_2$H$_5$PbI$_3$ and N$_2$H$_3$PbI$_3$ will maintain
a body-centered tetragonal structure, as seen in MAPbI$_3$.

Upon carrying out structure optimization, we find that the lattice constants for 
N$_2$H$_5$PbBr$_3$ and N$_2$H$_3$PbBr$_3$ are $a = 5.86 $ \AA \ and $5.806 $ \AA, \  respectively.
On the other hand, we find $a=8.81 $ \AA \  and $c=12.59 $ \AA \  for N$_2$H$_5$PbI$_3$, while
 $a=8.76 $ \AA$ $ and $c=12.52 $ \AA$ $ for N$_2$H$_3$PbI$_3$. The calculated band gaps of these
compounds, using the modified Becke-Johnson exchange potential with $A=0.4$ and $B=1.0$ bohr$^{1/2}$ in Eq.(3)
and taking into account spin-orbit coupling, are presented in Table 3. These results show that 
the replacement of CH$_3$NH$_3$ with N$_2$H$_5$ or N$_2$H$_3$ causes a considerable redshift
in the band gap values.

\begin{table*}[t]  
  \caption{ Band gaps for different lead halide compounds, calculated using the modified
Becke-Johnson exchange potential and taking into account the effect of spin-orbit coupling.}
      \begin{tabular}{cccccccc}
	\hline
	\hline 
          Compound      	& Band gap (eV)			\\
	\hline
	N$_2$H$_5$PbBr$_3$          & 1.94       		\\
	N$_2$H$_3$PbBr$_3$          & 1.77	 	  	\\
	N$_2$H$_5$PbI$_3$           & 1.41	 	  	\\
	N$_2$H$_3$PbI$_3$           & 1.13	 	  	\\
	\hline
	\hline
      \end{tabular}
      \label{tab:2}
\end{table*}

 \begin{figure}[htbp]
 \includegraphics[height=6.5cm]{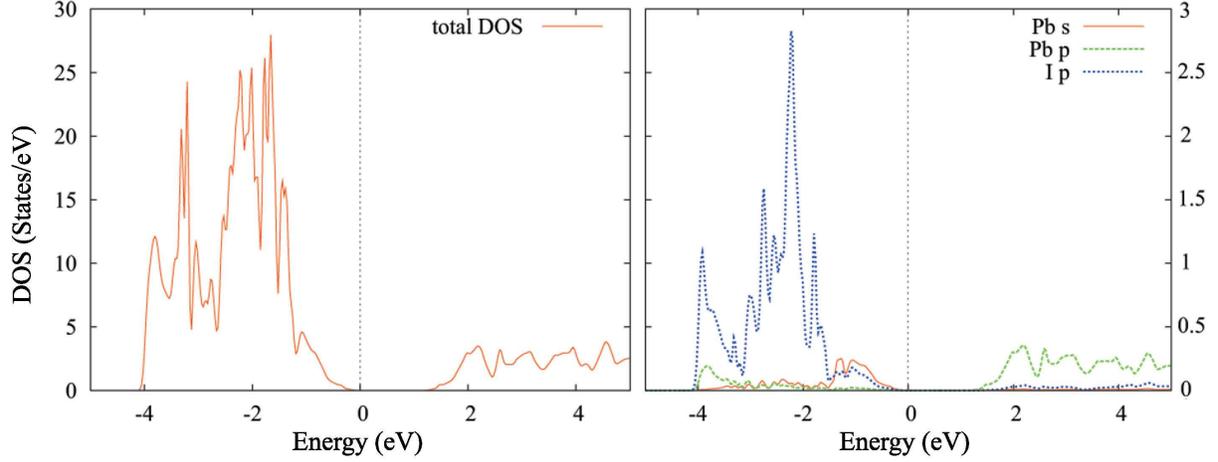}
  \caption{ Density of states (DOS) of N$_2$H$_5$PbI$_3$, obtained using the mBJ exchange potential and
 taking into account the effect of spin-orbit coupling.
  \label{fig:2}}
\end{figure}

The calculated density of states in N$_2$H$_5$PbI$_3$ is presented in Fig. 2. It is noted that,
similar to  CH$_3$NH$_3$PbI$_3$, the low-lying conduction bands are derived mainly
from Pb p orbitals, whereas the highest valence band is composed of both Pb s and I p states.
Bands in the energy range -4 eV to -2 eV are derived mostly from iodine p orbitals.
 The character of the valence and conduction bands
is also made clear in Fig. 3, where the energy bands along some high-symmetry directions
in the first Brillouin zone are plotted. 
The size of the circles is proportional to the contribution of the chosen atomic orbital
to the eigenstates at each \textbf{k}-point. The fact 
that s ($l=0$) and p ($l=1$) orbitals on the same atom (Pb) make large contributions to the 
wave functions at the valence band maximum and conduction band minimum is
responsible for the large optical absorption coefficients that occur in these compounds, and hence  
their usefulness in solar cell applications.

\begin{figure*}[t]
  \includegraphics[height=7.5cm]{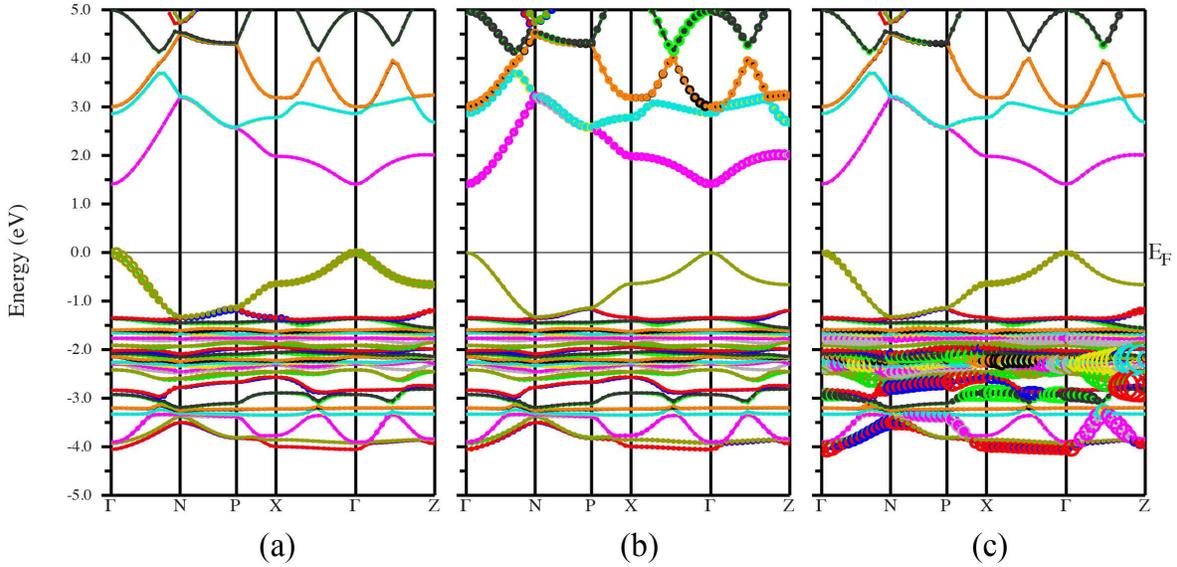}
  \caption{ Orbital character of the valence and conduction bands of N$_2$H$_5$PbI$_3$.
  The contribution of the selected orbital to the wave function (in a given band and at
a given \textbf{k}-point) is proportional to circle size, 
with a single point denoting zero contribution. (a) Pb 6s orbital, (b) Pb 6p orbital,
 and (c) I 5p orbital.
  \label{fig:3}}
\end{figure*}

 In  N$_2$H$_5$PbI$_3$, the valence band maximum (VBM)
and conduction band minimum (CBM) occur at the $\Gamma$-point, the Brillouin zone center. In ideal 
cubic perovskites, VBM and CBM occur at the zone's corner point R(1/2, 1/2, 1/2). Here, 
N$_2$H$_5$PbI$_3$ is assumed to have a  body-centered
tetragonal structure with two formula units per primitive cell. The conventional tetragonal
 unit cell, with four formula units, is a slight distortion of the   
 $\sqrt{2}\times\sqrt{2}\times2$ supercell of the ideal cubic unit cell, and point R is 
zone-folded into point $\Gamma$.

Spin-orbit coupling (SOC) has a profound effect on the band structure in organolead halide compounds.
In Fig. 3, we see that at the $\Gamma$ point, the lowest conduction band has energy 1.41 eV, 
while the next two higher bands have energy close to 3 eV.. In the absence of SOC, those three 
bands would be almost degenerate at the $\Gamma$ point, and all of them would occur at about 2.5 eV. 
In a cubic perovskite structure,
such as the one found in MAPbBr$_3$, the conduction band minimum at point R is six-fold
 degenerate (including spin degeneracy); SOC partially lifts the degeneracy, giving 
rise to a doublet ($j=1/2$) with a lower energy and a quartet ($j=3/2$) with a higher energy. In a 
body-centered tetragonal structure, CBM occurs at point $\Gamma$,  
degeneracy is now only approximate (it was exact in the cubic structure), and  SOC again splits
the almost six-fold degenerate level into one lower doublet and two higher doublets.

\section{Conclusions}\label{sec:4}

We have presented calculations on various lead halide compounds using density functional theory
with modified Becke-Johnson exchange potential. For the compounds PbCl$_2$, PbBr$_2$, and
CH$_3$NH$_3$PbBr$_{3-x}$Cl$_x$, for $x=0, 1, 2$, and $3$, we showed that the calculated band gaps
are in excellent agreement with experimental values. We then used this computational method to 
predict the electronic structure of similar compounds, namely, those that result from
the replacement of the methylammonium cation in MAPbBr$_3$ and MAPbI$_3$ with the slightly smaller
cations N$_2$H$_5^+$ and N$_2$H$_3^+$. A significant downshift in the band gap values is 
predicted to occur as a result of these replacements. In particular, we predict that N$_2$H$_5$PbI$_3$
and N$_2$H$_3$PbI$_3$ have band gaps given by 1.41 eV and 1.13 eV, respectively. Therefore. these compounds,
if synthesized, would be excellent light harvesters in solar cells. It should be noted, however, that the 
instability of the diazenium cation (N$_2$H$_3^+$) may make it difficult to use it as a replacement
for the methyl ammonium cation.

\acknowledgments

The author gratefully acknowledges support by National Science Foundation
 under grant No. HRD-0932421 and NSF PREM Program: Cal State L.A. \& Penn
State Partnership for Materials Research and Education, award DMR-1523588.

\section*{Conflict of interest}

The author reports no conflict of interest in this research.


\end{document}